\documentclass[journal,twoside,web]{ieeecolor}
\usepackage{tuffc}
\usepackage{cite}
\usepackage{amsmath,amssymb,amsfonts}
\usepackage{algorithmic}
\usepackage{graphicx}
\usepackage{textcomp}
\usepackage{wrapfig,colortbl}
\definecolor{abstractbg}{rgb}{1,0.969,0.914}
\def\BibTeX{{\rm B\kern-.05em{\sc i\kern-.025em b}\kern-.08em
    T\kern-.1667em\lower.7ex\hbox{E}\kern-.125emX}}
\begin{document}
\title{Orthogonal Linear Array based Product Beamforming for Real Time Underwater 3D Acoustical Imaging}
\author{ Mimisha M Menakath, \IEEEmembership{Student Member, IEEE}, Mahesh Raveendranatha Panicker,  \IEEEmembership{Senior Member, IEEE} and Hareesh G, Jr., \IEEEmembership{Member, IEEE}
\thanks{``This work was supported by Ministry of Education as a part of DRDO-MHRD Research Scheme.'' }
\thanks{ Mimisha M Menakath is with the Electrical Engineering Department, Indian Institute of Technology Palakkad (e-mail: 122014006@smail.iitpkd.ac.in).}
\thanks{Mahesh Raveendranatha Panicker, was with Indian Institute of Technology, Palakkad. He is now with the Infocomm Technology Cluster, Singapore Institute of Technology (e-mail:mahesh.panicker@singaporetech.edu.sg).}
\thanks{Hareesh G is with Naval Physical Oceanographic Laboratory, Kochi (e-mail:hareeshg.npol@gov.in, rubinpeter.npol@gov.in).}}

\IEEEtitleabstractindextext{%
\fcolorbox{abstractbg}{abstractbg}{%
\begin{minipage}{\textwidth}\rightskip2em\leftskip\rightskip\bigskip
\begin{wrapfigure}[15]{r}{3in}%
\hspace{-3pc}\includegraphics[width=2.9in]{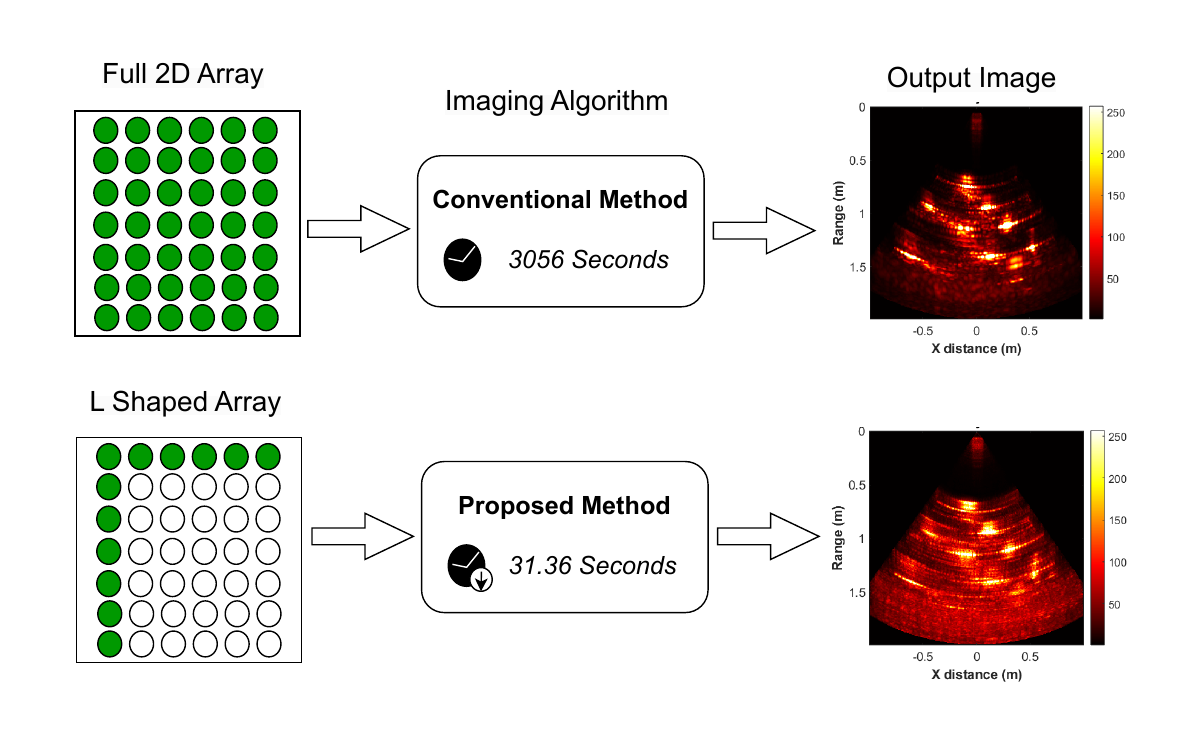}
\end{wrapfigure}%
\begin{abstract}
Ocean exploration using acoustical 3D imaging is gaining popularity as it provides information about the 3D geometry of the targets even under mild turbid conditions. A major challenge in underwater 3D imaging is the high cost of the planar arrays and the computational complexity of the image reconstruction algorithms. In this work, we introduce two novel aspects, an L shaped array and a quadrant based time domain receive beamforming with a focus on achieving relatively low computational complexity for real-time 3D imaging in underwater environments. An orthogonal combination of two linear arrays to form an L shape is used to perform two independent and parallel 2D delay and sum beamforming and the 3D image of the target is reconstructed using the product of the resulting beams. In the proposed quadrant-based beamforming, each quadrant in the imaging slice is reconstructed in parallel using the orthogonal L-shaped linear arrays which reside at the edges of the planar array for the quadrant to be reconstructed. The proposed L-array solves the multiple-target ambiguity issue of the cross-array and an L-array placed at the center of the uniform planar array. The proposed method has reduced the main lobe width by $1^\circ$ with $11.8$ $dB$ increase in the peak side lobe level when compared to the conventional delay and sum beamforming using a uniform planar array. Although there is an increase in the side lobe level, the asymmetric beam pattern of the proposed L array and the quadrant-based beamforming restrict the side lobes within a quadrant. The proposed method achieves a reduction in computation time by a factor of 97 for 3D imaging compared to the conventional method, while maintaining acceptable image quality. For qualitative analysis, the 3D images of different underwater targets have been reconstructed and compared in simulation and experiment.
\end{abstract}

\bigskip
\end{minipage}}}

\maketitle

\section{Introduction}
\label{introduction}

Underwater 3D imaging has been gaining significance as a promising technique for ocean exploration. In the past, 3D sonars were used primarily for pipeline monitoring, structural inspections, and trench surveys under highly turbid conditions. Its ability to capture three-dimensional information about underwater targets now makes them useful for obstacle avoidance in autonomous underwater vehicles (AUVs). However, the requirement of a uniform planar array with a large number of sensors restricts the real-time performance of the system.

Many sparse array design techniques \cite{zhao2021pruned, zhao2018optimized, gu2020optimization, chen2010optimized, chi2016ultrawideband} have been proposed to overcome the limitation. In the sparse array-based methods, the number of sensor elements is optimally reduced and distributed over the same aperture as the uniform planar array to maintain the similar image resolution. However, the sparse array retains a relatively high number of elements, and real-time 3D beamforming continues to be challenging. The Mills cross-array has been used for underwater 3D acoustic imaging as an alternative to the uniform planar array \cite{7003133,7156166}. In the Mills-cross array, one linear array handles transmission, while the other one handles the reception. It requires multiple signal transmissions and scans, which limits its ability to perform in real time.   In addition to cross-arrays, different combinations of horizontal and vertical linear arrays have also been explored for 2D DOA estimation in radio astronomy and RADAR \cite{slattery1966use,tayem2005shape}. 
In RADAR and radio astronomy, a spherical coordinate system is employed that is distinct from that used in underwater imaging. Furthermore, the L-shaped and cross-shaped receiving arrays with product beamforming lead to multiple-target ambiguity issues as explained in \cite{slattery1966use}. To the best of our knowledge, the literature on the use of orthogonal linear arrays, other than the cross-array, for underwater 3D acoustic imaging is very limited.

 On the receive signal processing, a few fast beamforming algorithms have been proposed to decrease the computational complexity associated with traditional 3D beamforming. In \cite{murino1994underwater}, Murino et al. suggested combining dynamic focusing with fast Fourier transform (FFT) beamforming for short-range imaging using narrowband signals. A distributed and parallel subarray (DPS) beamforming was proposed for narrowband transmission by Han et al in \cite{han2013real}. A similar approach has been proposed in \cite{zhao2018efficient}, where Zhao et al. utilized a strategy in which they divided the linear receiving array of the cross array into two-stage subarrays. In \cite{rypkema2020memory}, Rypkema et al. introduced an approach for approximate 3D beamforming applications involving small arrays. Recently, Zhao et al. introduced the concept of fine grid sparse arrays along with a new beamforming algorithm for a narrowband underwater imaging system in \cite{zhao2021pruned}. All the aforementioned beamforming algorithms are performed in the frequency domain and they are extensively used for narrowband transmission. Two beamforming algorithms for wideband transmission were described in \cite{chi2015fast, palmese2010efficient}. In \cite{chi2015fast}, Chi et al. proposed an efficient, but approximate method for 3D underwater acoustic imaging with wideband transmission. In \cite{palmese2010efficient}, Palmese et al. proposed an accurate 3D beamforming method for wideband transmission based on the chirp zeta transform (CZT).
 
 The fast beamforming algorithms and sparse array techniques documented in the literature effectively reduce the computation time required for 3D beamforming and facilitate the generation of 3D images in near real-time. This efficiency may occasionally result in a trade-off with image resolution. However, the real-time performance of these algorithms for critical applications, such as obstacle avoidance, requires further reductions in computation time and array complexity. In this work, we utilize a novel 2-D L-shaped linear array that features a very limited number of sensor elements, combined with a nonlinear beamforming, to create real-time 3D underwater images. The method is employed in the time domain ensuring that the algorithm's complexity remains consistent regardless of the transmitted signal bandwidth.
\begin{table*}[!t]
\arrayrulecolor{subsectioncolor}
\setlength{\arrayrulewidth}{1pt}
{\sffamily\bfseries\begin{tabular}{lp{6.75in}}\hline
\rowcolor{abstractbg}\multicolumn{2}{l}{\color{subsectioncolor}{\itshape
Highlights}{\Huge\strut}}\\
\rowcolor{abstractbg}$\bullet$ & A novel L shaped array combined with a nonlinear product beamforming is proposed for real time underwater 3D acoustical imaging.\\
\rowcolor{abstractbg}$\bullet${\large\strut} & The impact of increased side lobes from reducing the number of sensors is minimized using a quadrant-based approach. \\
\rowcolor{abstractbg}$\bullet${\large\strut} & The proposed method achieves a reduction in computation time by a factor of 97 for 3D imaging compared to conventional method, while maintaining acceptable image quality. \\[2em]\hline
\end{tabular}}
\setlength{\arrayrulewidth}{0.4pt}
\arrayrulecolor{black}
\end{table*}
\subsection{Motivation and Contributions}
The proposed work aims to conduct further research in two broad areas of underwater 3D acoustical imaging, as described in the following. \\ 
\subsubsection{Efficient Array Design} A uniform planar array is ideal for 3D imaging; however, the design and computational cost have motivated researchers to investigate linear array combinations and sparse arrays as alternative options. In this work, we have presented the use of a uniform rectangle array at the edges of the planar array instead of a uniform planar array for far-field 3D imaging. During reconstruction, only an L-shaped configuration is employed to reconstruct each of the quadrants. This modification not only reduces hardware complexity, but also minimizes memory requirements.\\
\subsubsection{Non-linear product beamforming} Traditionally, narrowband transmission has been used for underwater imaging. Thus, most of the fast-beamforming algorithms have been proposed in the frequency domain. The primary obstacle preventing the use of time-domain beamforming for underwater 3D imaging was the substantial computational complexity arising from the hardware interpolation filter \cite{chi2019underwater}. In scenarios where wideband transmission is required for imaging with high resolution \cite{chi2017high}, and there is a prevalence of software interpolation using multi core computing systems, the computation time of the direct method in the frequency domain is anticipated to exceed that of time domain beamforming \cite{menakath2022comparison}. 
In this paper, a new and efficient nonlinear beamforming method in the time domain is introduced for 3D acoustical imaging for real-time underwater applications. The proposed algorithm improves image resolution and minimizes the effects of side lobes that arise from the reduction of sensor elements. \\

The structure of the paper is as follows. Three state-of-the-art methods for receiving beamforming are briefly described in Section II. In Section III, the proposed L-shaped orthogonal linear array-based product beamforming is explained with a detailed analysis of the point spread function and computation complexity. Section IV explains underwater simulation using the k-wave toolbox with the required parameter definitions. Sections V and VI discuss the simulation results and the experimental results, respectively. The paper concludes with Section VII.

\section{Background}
The essential elements of an underwater acoustical 3D imaging system comprise an omnidirectional projector and a uniform 2D receiving array. The received backscattered echoes are processed using beamforming to convert the received time information into space information in the form of an image. A brief overview of typical receive beamforming approaches is detailed in the following subsections.
\subsection{Time domain Delay and Sum Beamforming (DAS)}
Delay and sum beamforming is the conventional technique employed for image reconstruction in underwater acoustical 3D imaging \cite{murino2000three}. In DAS beamforming, the receive uniform 2D array concentrates on every voxel outlined in the imaging grid. This is achieved by appropriately delaying the signals received by each sensor element and summing up these delayed signals to determine the beam corresponding to the voxel. The side lobes of the array can be minimized by applying a weighting technique called shading. In this work, we used equal weights (rectangular window shading) for all sensors.
\subsection{Frequency domain Direct Method (DM)}
The delay and sum beamforming can also be performed in the frequency domain. This involves applying delay and sum operations to signals in the frequency domain, offering an alternative approach to the traditional implementation in the time domain \cite{chi2019underwater}. In DM, the received signal is partitioned into distinct overlapping blocks of length L to achieve range (time) localization of the Discrete Fourier Transform (DFT). The optimal number of samples to be overlapped for computing all the beam signal samples has been thoroughly investigated in \cite{palmese2010efficient}.
\begin{figure*}
\centerline{\includegraphics[width=0.7\textwidth]{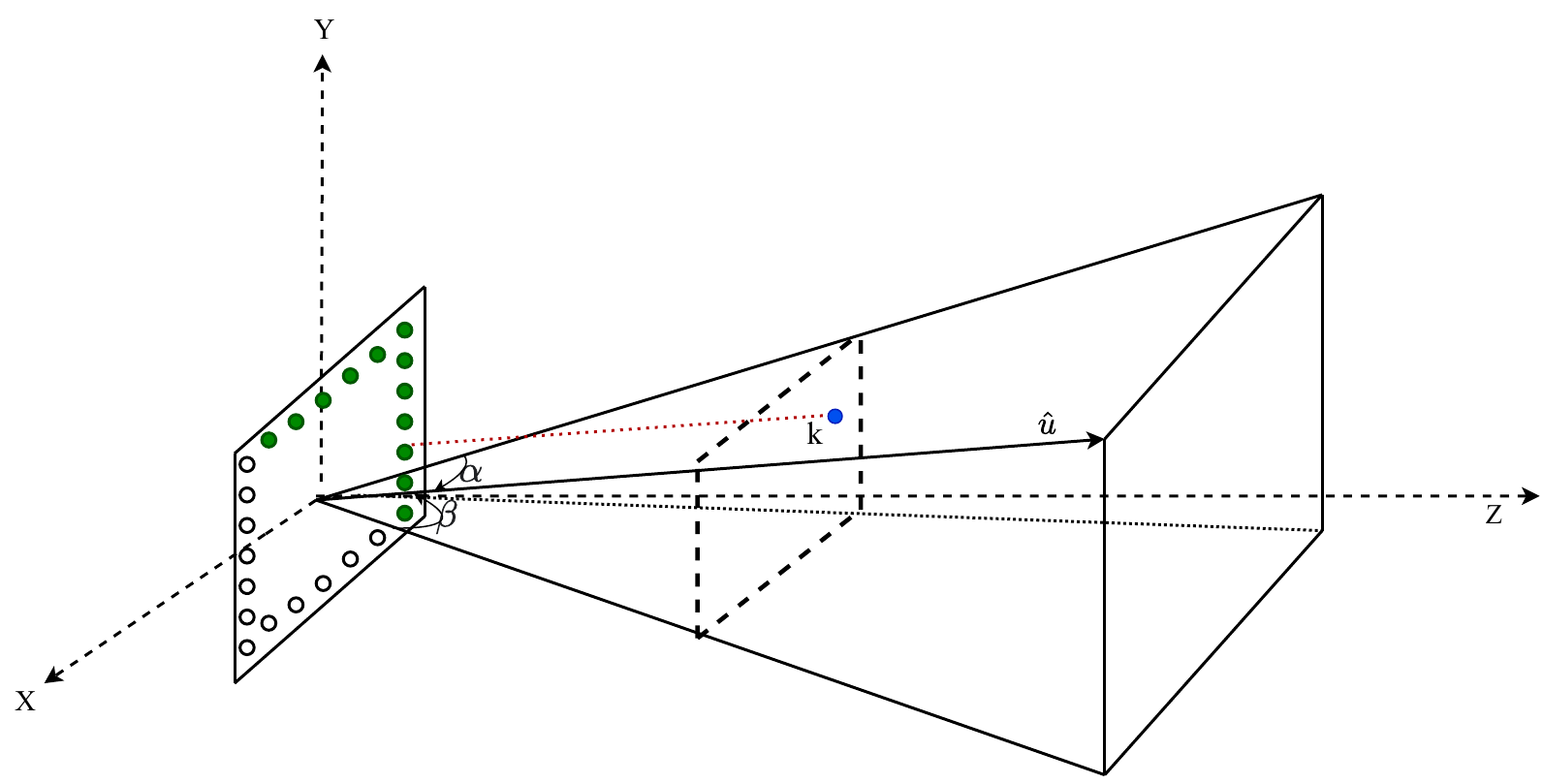}}
\caption{ Coordinate geometry of the imaging grid with a rectangle receiving array. $\alpha$ and $\beta$ represent azimuth and elevation angle respectively for the direction, $\hat{u}$. The blue dot ($k$) represents a pixel on the image slice represented as a rectangle with dash lines. The selected L array for the pixel, $k$ is highlighted in green color.}
\label{fig1}
\end{figure*}
\subsection{Chirp Zeta Transform Beamforming (CZT)}
Chirp zeta transform beamforming \cite{palmese2007chirp} is a fast implementation technique for delay and sum beamforming in the frequency domain utilizing the chirp-zeta transform (CZT). In CZT beamforming, the mathematical equation of DFT of the beam in a particular direction is modified using the CZT transform. The modified equation can be easily implemented by performing the 2D discrete convolution of two functions as explained in \cite{palmese2007chirp}. The 2D-discrete convolution is performed by using 2D-FFT which accelerates the operation. Since the method gives all the beams in a slice together, the method is fast. Also, the method gives the same result as the time-domain delay and sum beamforming
gives. This method offers an excellent balance between speed and image quality, surpassing other fast beamforming techniques in the frequency domain that have been proposed for wideband signals.

\section{The Proposed Methodology}
\subsection{L-shaped Orthogonal Linear Array with Product Beamforming}

\begin{figure*}
\centering
\includegraphics[width=\textwidth]{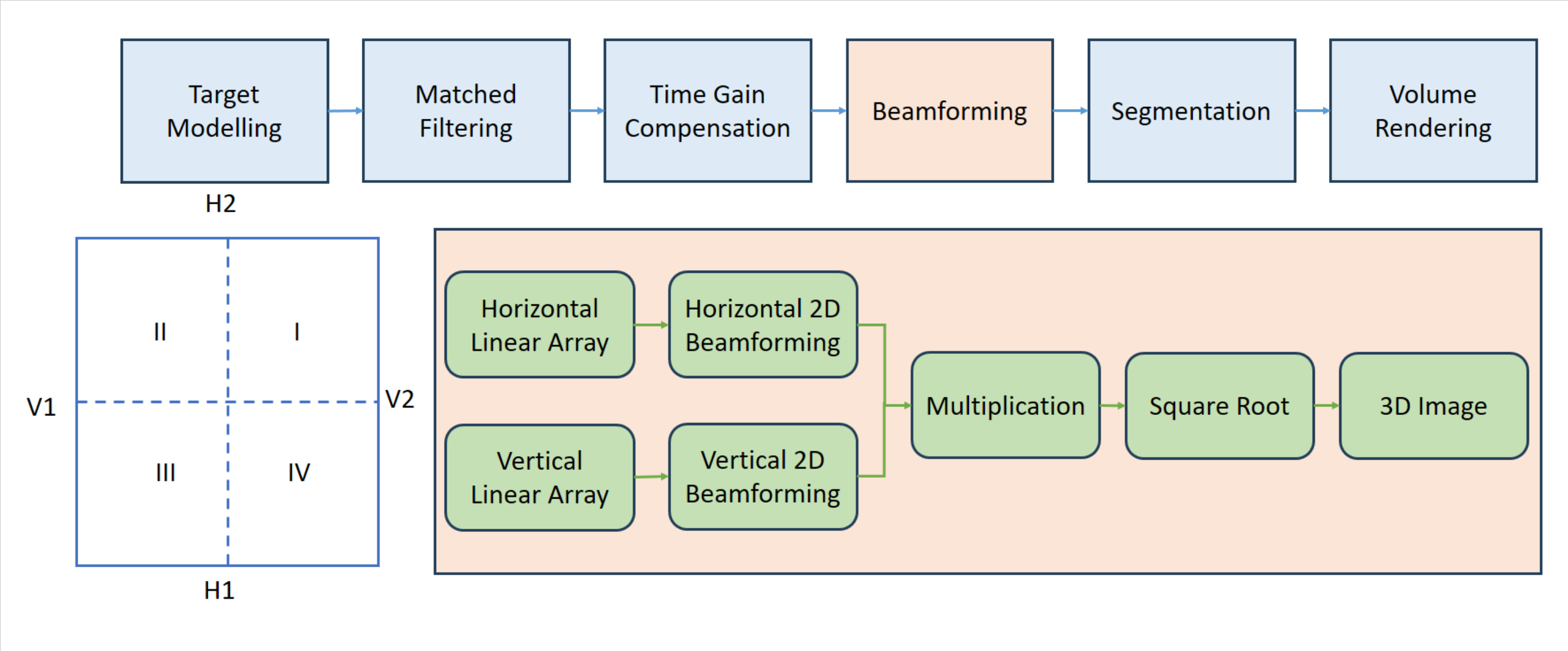}
\caption{The block diagram of the proposed 3D image reconstruction and segmentation workflow with emphasis on beamforming. The selection of the orthogonal linear arrays for the proposed quadrant based beamforming is also illustrated.  }
\label{fig2}
\end{figure*}
The proposed method uses a rectangle array with four linear arrays placed at the periphery of the conventional uniform planar array to reconstruct the entire 3D space. The coordinate geometry of the imaging grid is shown in Fig.\ref{fig1}. A two-dimensional receiving rectangle array with $2M+2N$ omnidirectional and point sensors is located in the plane, $Z=0$. In the proposed method, two independent 2D beamforming is performed instead of a single 3D beamforming, by using an orthogonal combination of linear arrays, i.e. an L-shaped array. The L array is chosen according to the quadrant in which the pixel to be reconstructed is situated, as illustrated in Fig.\ref{fig1}. Consider $(m,n)$ as the index of the sensor element in the uniform planar array of elements $M\times N$ and $(m_{sel},n_{sel})$ as the index of the common sensor element of the orthogonal linear arrays. The horizontal and vertical beamforming is performed as in \eqref{1} and \eqref{2}, respectively.  
\begin{multline}
B_H\left(r_0, \alpha_p, \beta_q, t\right)\\
= \sum_{m=1}^{M} W_{m, n_{sel }} S_{m, n_{sel }}(t 
\left.-\tau\left(m, n_{sel }, r_0, \alpha_p, \beta_q\right)\right) .
\label{1}
\end{multline}
\begin{multline}
B_V\left(r_0, \alpha_p, \beta_q, t\right)\\  =\sum_{n=1}^N W_{m_{sel , n}} S_{m_{sel }, n}(t \left.-\tau\left(m_{sel }, n, r_0, \alpha_p, \beta_q\right)\right) .
\label{2}
\end{multline}

$S_{m,n } (t)$ denotes the signal received for the sensor element indexed by $(m,n)$ and $W_{m,n}$ signifies the shading weight associated with the respective sensor element. The unit vector towards the direction of  $(\alpha, \beta)$, designated as $\hat{u}$, is described in equation \eqref{4},
\begin{equation}
\widehat{\mathrm{u}}=\left(\sin \alpha, \sin \beta, \sqrt{\cos ^2 \alpha-\sin ^2 \beta}\right),
\label{4}
\end{equation}
where, $\alpha$ is the azimuth angle and the $\beta$ is elevation angle as depicted in Fig.\ref{fig1}.

The delay needed to direct the beam towards the scatterer at  $(r_0, \alpha_p,\beta_q )$ for the sensor element at $(x_m,y_n,0)$ can be expressed as \cite{wang2021improving}
\begin{multline}
\tau\left(m, n, r_0, \alpha_p, \beta_q\right)=\\
\frac{r_0-\sqrt{\begin{array}{c}
\left(r_0{ }^2+x_m{ }^2+y_n{ }^2-2 x_m r_0 \sin \alpha_p\right.
\left.-2 y_n r_0 \sin \beta_q\right)
\end{array}}}{c},
\label{5}
\end{multline}
where $r_0$ is the focusing distance.
The condition for the far-field approximation is given by $r_0>\frac{D^2}{2\lambda}$, \cite{wang2021improving}, where $D$ represents the aperture size of the array and $\lambda$ is the wavelength. The far-field delay can be approximated as \eqref{6} \cite{palmese2009three}
\begin{equation}
\tau\left(m, n, \alpha_p, \beta_q\right)=\frac{x_m \sin \alpha_p+y_n \sin \beta_q}{c},
\label{6}
\end{equation}
where, c is the speed of sound in the medium. 

The final time domain beamformed signal, $B_k (r_0,\alpha_p,\beta_q,t)$ for the $k^{th}$ scatterer at a focusing distance $r_0$ in the steering direction $(\alpha_p,\beta_q )$  is acquired by multiplying the results of the horizontal and vertical beamformed values, as described in \eqref{3}. The square root of the product is taken to make it the same quantity as the delay and sum beamforming using a uniform planar array. The proposed and conventional beamformed signals are scaled by $1/(M\times N)$ for a normalized comparison.

\begin{equation}
\begin{aligned}
& B_k\left(r_0, \alpha_p, \beta_q, t\right) \\
& \quad=\frac{1}{M N} \sqrt{\left(B_H\left(r_0, \alpha_p, \beta_q, t\right) \times B_V\left(r_0, \alpha_p, \beta_q, t\right)\right.} \\
& =\frac{1}{M N} \sqrt{\sum_{m=1}^M \sum_{n=1}^N\left(W_{m, n_{s e l}} W_{m_{s e l, n}} S_{m, n_{s e l}}\left(t-\tau_{m, n_{s e l}}\right)\right.} \\
& \left.\quad S_{m_{s e l, n}}\left(t-\tau_{m_{s e l}, n}\right)\right),
\end{aligned}
\end{equation}

where p varies from $0$ to $M_b$ (discrete $\alpha$) and $q$ varies from $0$ to $N_b$ (discrete $\beta$). $M_b$ is the total number of beams in the azimuth direction and $N_b$ is the total number of beams in the elevation direction. The 3D image is obtained by stacking 2D slices formed by $M_b\times N_b$ beams at different ranges.  
 
Mathematically, the proposed method is the cross-correlation of the delay-compensated signal of one linear array to that of its orthogonal linear array. So, the method will boost the target echo while reducing clutter in the background due to the cross-correlation operation as in \cite{matrone2014delay} and \cite{8485325}. Real-time performance can be achieved by performing vertical and horizontal beamforming in parallel. 
\begin{figure*}
\centering
\includegraphics[width=0.7\textwidth]{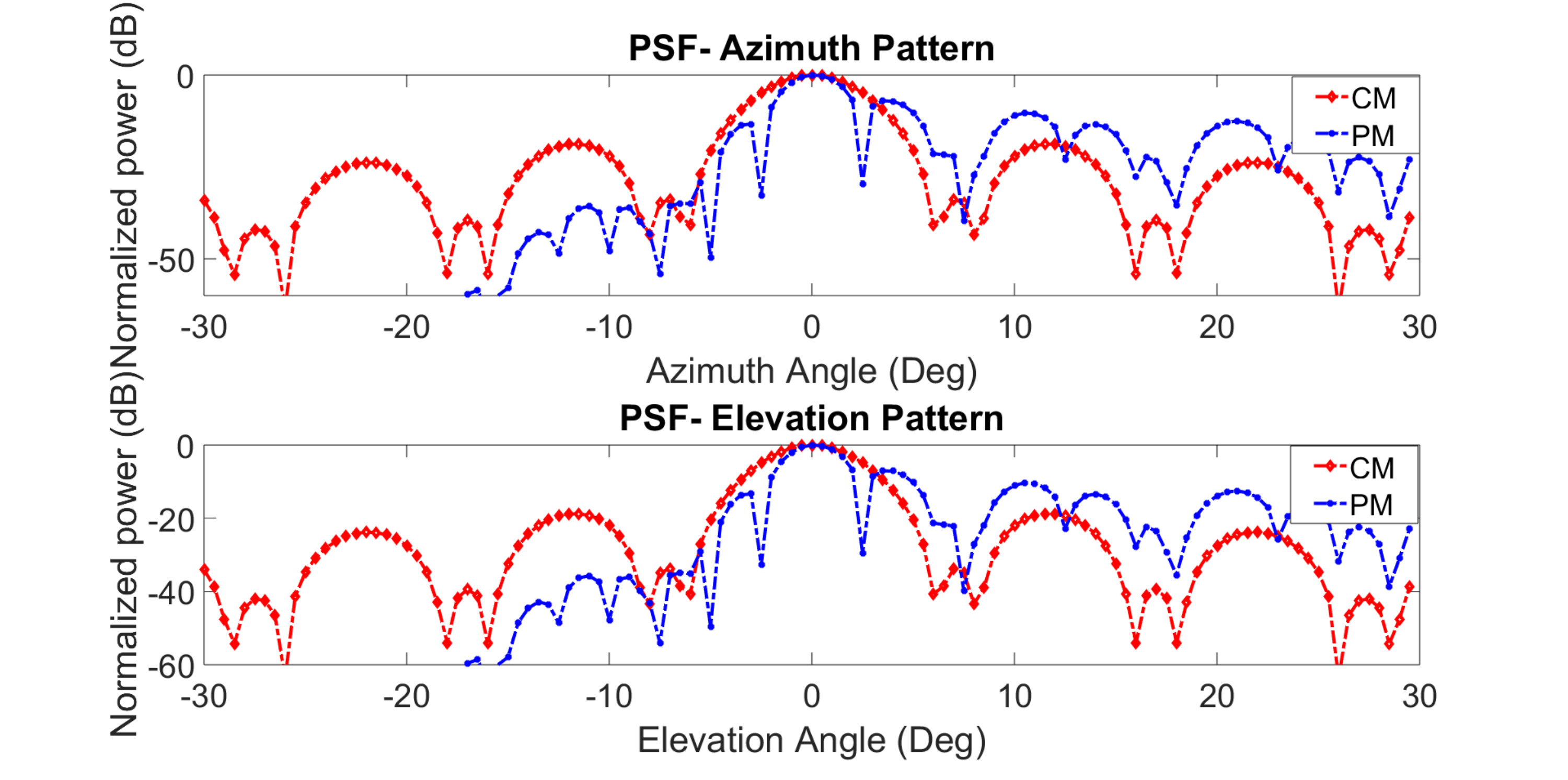}
\caption{Azimuth and elevation pattern of the point spread function of the proposed method (PM) in which proposed product beamforming is performed with L-shaped array of 47 elements and the conventional method (CM) in which delay and sum beamforming is proposed with a uniform rectangular array of 576 elements.}
\label{fig3}
\end{figure*}
The schematic representation of the 3D image reconstruction workflow is shown in Fig.\ref{fig2}. The backscattered echoes received at the sensors are matched filtered to increase signal-to-noise ratio (SNR). As the impulse response of the matched filter is the conjugated time-reversed form
of the transmitted signal, it will correlate with the received signal and results in a high value at the time instant when the received signal completely matches with the transmitted signal and low values at the
rest of the time instants. So, it reduces noise in the received signal while enhancing the known signal and thus improves the signal-to-noise ratio \cite{1454543}. The matched filtered signal is then fed to the time gain compensator (TGC) in order to compensate the attenuation due to depth. Following matched filtering and Time Gain Compensation (TGC), the received backscattered echoes are employed to reconstruct the 3D image of the underwater target. This reconstruction is accomplished by using the proposed orthogonal linear array product beamforming algorithm in the time domain.

In the proposed 3D image reconstruction method, the imaging plane at a particular range containing $M_b\times N_b$ beams can be divided into four quadrants (I, II, III and IV) as shown in Fig.\ref{fig2}. Horizontal linear arrays are represented as H1 and H2, while vertical linear arrays are represented as V1 and V2.  Thus, linear arrays can be chosen as those which form the edges of the quadrant where the pixel resides. For example, linear arrays labeled as H2 and V2 are used to reconstruct quadrant I and so on.  The pixels in $(0^\circ,0^\circ)$ can be reconstructed by reconstructing all quadrants together. This method replaced the requirement of a computationally complex uniform planar array with a low-complex rectangle array that has only four linear arrays at the periphery of the planar array. Thus, the total number of sensors required in the uniform planar array is reduced from $MN$ to $2(M+N)$. Also, if all quadrants are reconstructed parallel, it will take only one fourth of the time required for the entire slice reconstruction. It makes the implementation easy and allows us to further reduce the computation time. The proposed L-shaped array has advantages of lower main lobe width (at the expense of higher side lobes) and asymmetry in point spread function, which helps in unambiguous quadrant reconstruction, which are thoroughly analyzed in the subsequent section.

For better 3D image representation, the reconstructed 3D volume is segmented using k-means clustering \cite{kanungo2002efficient} and converted to a binary image by taking a cluster having the maximum mean intensity. The segmented binary image is then scan-converted. Since the image is reconstructed in a polar coordinate system, conversion to a Cartesian coordinate system is necessary to make it suitable for display, since most display devices are designed for the Cartesian coordinate system \cite{szabo2004diagnostic}. The scan conversion is performed in three steps. 1) defining a new imaging grid in Cartesian, 2) Cartesian to polar conversion of the new imaging grid, and 3) trilinear interpolation. After scan conversion, 3D volume rendering is achieved by using the function \textit{voxelPlot} in the k-Wave toolbox.

\subsection{Analysis of Point Spread Function}
A thorough analysis of the proposed L-shaped array and the orthogonal linear array product beamforming is explained in this section. The performance of the proposed method is compared with that of the conventional method in which DAS beamforming is employed with a uniform rectangular array (URA). The azimuth and elevation pattern of the point spread function (PSF) is plotted as shown in Fig. \ref{fig3}. As the main lobe width (MLW) at $-3dB$ power of the proposed method is $1^\circ$ lower than that of conventional method, the resolution of the image has been improved by $1^\circ$. However, the peak side lobe level (PSLL) has increased by $11.8$ $dB$ due to the reduction  of number of sensors, increasing the background noise. Since for the proposed method using an L array, the side lobes are significant only on one side of the point as shown in Fig.\ref{fig3}, we used different L arrays to reconstruct different quadrants in the imaging plane. So, the background noise due to the side lobes is confined only to the quadrant where the point is placed.

\begin{figure*}
\centering
\includegraphics[width=0.7\textwidth]{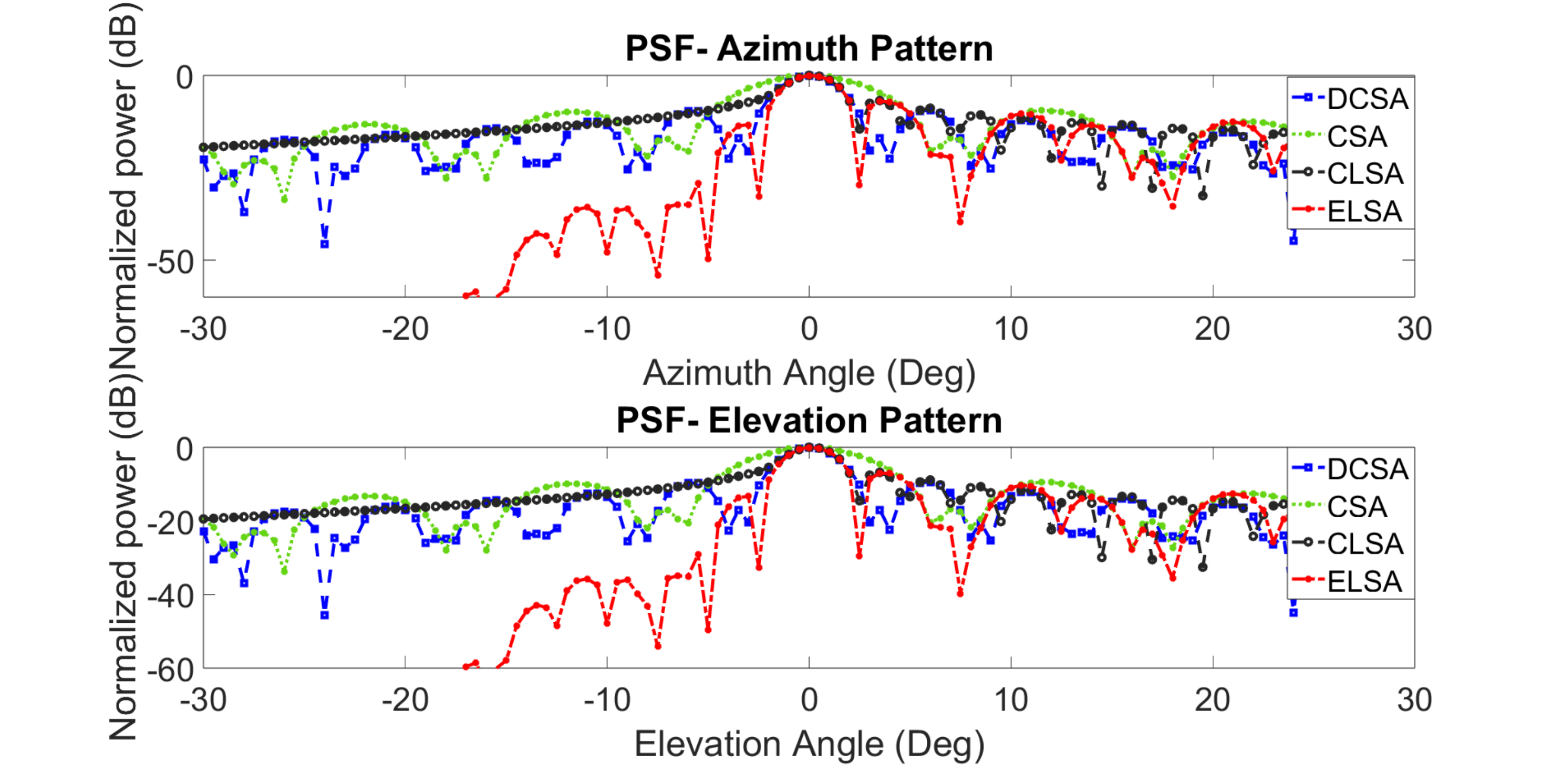}
\caption{Comparison of azimuth and elevation pattern of the point spread function of the proposed L-shaped array at the edges (ELSA), the L shaped array at the center (CLSA),  the cross shaped array (CSA), and  the double length cross array (DCSA) for the product beamforming.}
\label{fig4}
\end{figure*}

\begin{table}
\centering
\caption{Image Quality Analysis}
\label{table1}
\setlength{\tabcolsep}{5pt}
\begin{tabular}{p{35pt}p{30pt}p{30pt}p{35pt}p{35pt}p{30pt}}
\hline\hline
Array Geometry& 
Number of Elements & 
Method&
Azimuth MLW (Deg)&
Elevation MLW (Deg)&
PSLL (dB)
\\
\hline
URA& 
576& 
CM&
3& 
3& 
-18.8\\

ELSA& 
47& 
PM&
2& 
2& 
-7\\
CLSA& 
47& 
PM&
2& 
2& 
-7\\
CSA& 
47& 
PM&
5& 
5& 
-9.5\\
DCSA& 
95& 
PM&
2& 
2& 
-7\\
\hline\hline
\end{tabular}
\label{tab1}
\end{table}

For further analysis, the proposed L-shaped array at the edges (ELSA) is compared with the L-shaped array at the center (CLSA) (i.e., the intersection of two sub-arrays meets the origin), the cross-shaped array (CSA) and the double-length cross-array (DCSA) having twice the number of elements in each sub-array. The azimuth and elevation pattern of the PSF of the arrays for the proposed product beamforming is plotted as shown in Fig. \ref{fig4} and compared in TABLE \ref{table1}.  For ELSA, even though the MLW is reduced by $3^\circ$ as compared to CSA, the PSLL has increased by $2.5$ $dB$.  Since the PSF of CSA is symmetric, there is an issue of ambiguity as explained in literature \cite{slattery1966use}.  However, the asymmetric nature of ELSA solves the problem of ambiguity. This is because of the phase change of the beam pattern of each linear array of ELSA in its orthogonal direction. The mathematical derivation of the beam pattern of the ELSA and the simulation results of a detailed comparison of ELSA and the CSA for the product beamforming are provided as the supplementary material. The asymmetry in the PSF is obtained not only by the shape of the array but also by the geometric position of the array. To prove this, we have compared the PSF patterns of ELSA and CLSA. It was observed that the PSF pattern of the CLSA is not exactly symmetric, but the side lobe level is almost the same on both sides of the point. So, the array shows ambiguity, as explained in \cite{tayem2005shape}. The ELSA performs as similar as a DCSA in terms of MLW and PSLL. Also, the proposed ELSA solves the ambiguity issue of the DCSA as explained previously. The increase in the number of sensor elements by twice increases the computational complexity and hardware complexity by twice for a DCSA as compared to the ELSA. Thus, ELSA performs better than all other orthogonal array configurations considered.

\begin{figure*}
\centering
\includegraphics[width=\textwidth]{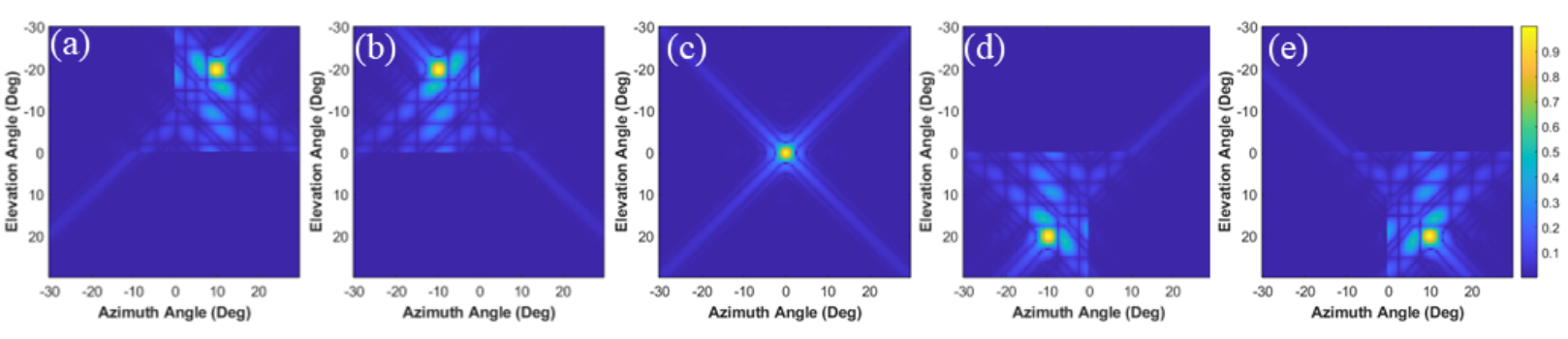}
\caption{Images of different point sources in different quadrants and at the center using the orthogonal linear array product beamforming.}
\label{fig5}
\end{figure*}

To validate the proposed method, different point sources are modeled at the $30$ $m$ range at different azimuth and elevation angles. A wideband continuous wave with center frequency of $500$ $kHz$ and bandwidth of $200$ $kHz$ is transmitted. The back scattered echoes are acquired by a uniform planar array of $24\times24$ elements with half-wavelength element spacing. The images were reconstructed by orthogonal linear array beamforming using the proposed ELSA. The obtained images are shown in Fig.\ref{fig5}. The proposed method could accurately reconstruct the images of the point sources in the modeled position in all quadrants. The reconstructed points are more confined to their position due to reduction in the MLW. However, as discussed earlier, the proposed method produces background noise as a result of side lobes in the quadrant where it is placed.

\begin{figure}
\centerline{\includegraphics[width=\columnwidth]{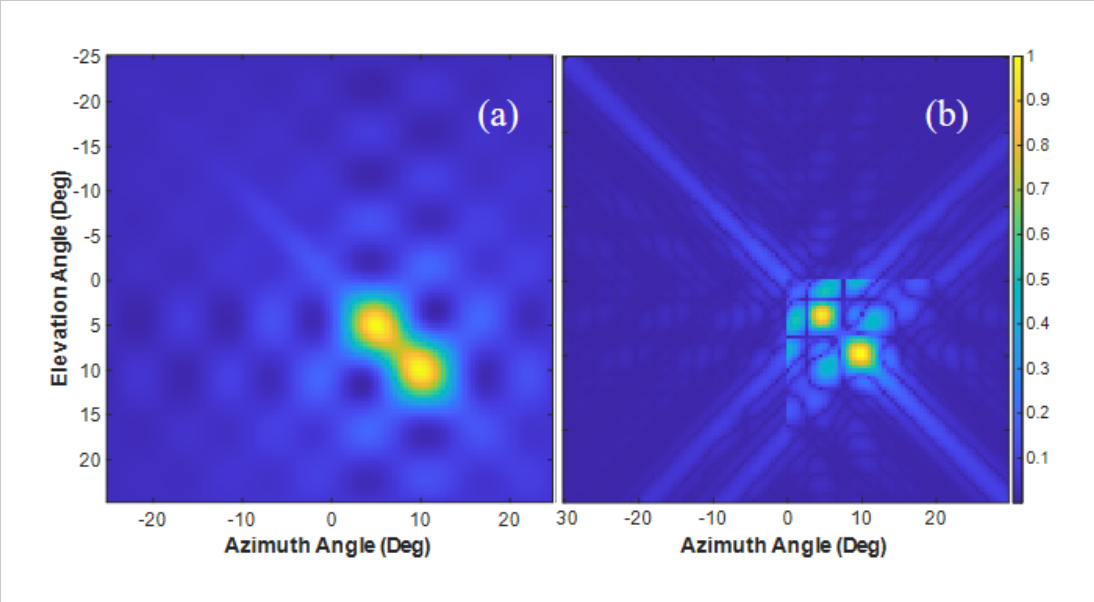}}
\caption{Images of two nearby points to analyse resolution of different methods. (a) Conventional method. (b) Proposed method.}
\label{fig6}
\end{figure}

Since the angular resolution is improved by the proposed method, the method is able to distinguish two nearby points, while the conventional method struggles to distinguish them, as shown in Fig.\ref{fig6}. Two point sources are placed at the angles $(5^\circ, 5^\circ)$ and $(10^\circ, 10^\circ)$ at the $30$ $m$ range. The images reconstructed by the conventional and the proposed methods are shown in Fig.\ref{fig6}(a) and (b) respectively. The two point sources are not resolvable in Fig.\ref{fig6}(a), while it is clearly resolved as two separate point sources in Fig.\ref{fig6}(b).

\subsection{Computational Complexity}
The number of real operations required for $N_b^2$ beams with a square array of $N^2$elements for time domain beamforming, frequency domain beamforming, CZT beamforming and the orthogonal linear array beamforming are compared in this section.

\textit{Time Domain beamforming:} The quantity of addition and multiplication operations needed for the execution of traditional time-domain beamforming are  $N_b^2 (N^2-1)$ and $N_b^2 N^2$ respectively \cite{chi2019underwater}. An interpolation operation is performed to find the values corresponding to the delays. If linear interpolation is utilized to estimate the values of sensor data for the required delays, the amount of additions and multiplications required would be $2N_b^2 N^2 $and $N_b^2 N^2$ respectively. 

\textit{Frequency Domain Beamforming:} The count of real operations needed to determine the DFT coefficient of the beam for an individual frequency bin is documented in \cite{chi2019underwater} as $(8N^2-2) N_b^2$. For $L$ frequency bins, the number of real operations required is $L(8N^2-2) N_b^2$. The number of operations required for the initial FFT to convert an $L$ length time domain signal into the frequency domain is $(2.5Llog_2  L/2+7L) N^2$ and final IFFT to convert the DFT coefficients of beams in the frequency domain into the time domain requires $5N_b^2 Llog_2 L$ operations.

\textit{Chirp Zeta Transform Beamforming:} The count of real operations needed to calculate the DFT coefficient of the beam corresponding to a single frequency is specified in \cite{chi2019underwater} as $ 6×[N^2+N_b^2+L^2 ]+20L^2 log_2 (L)$. For L frequency bins, it is $ 6 L×[N^2+N_b^2+L^2 ]+20L^3 log_2 (L)$. As in frequency domain direct method, the number of operations required for initial FFT and final IFFT are $(2.5Llog_2  L/2+7L) N^2$ and $ 5N_b^2 Llog_2 L$ respectively.

\textit{Orthogonal Linear Array Product Beamforming:} Since the proposed method uses linear arrays for beamforming, the number of additions required for vertical and horizontal beamforming is $ 2N_b^2×(N-1)$. The quantity of multiplications needed for vertical and horizontal beamforming is $2N_b^2 N $. It has to be noted that additional $N_b^2$ multiplications and square root operations are required for finding the final beam values. Given that linear interpolation is employed to estimate the delayed signal, the required number of multiplications and additions is $2N N_b^2$ and $4N N_b^2$, respectively.
\begin{figure*}
\centering
\includegraphics[width=\textwidth]{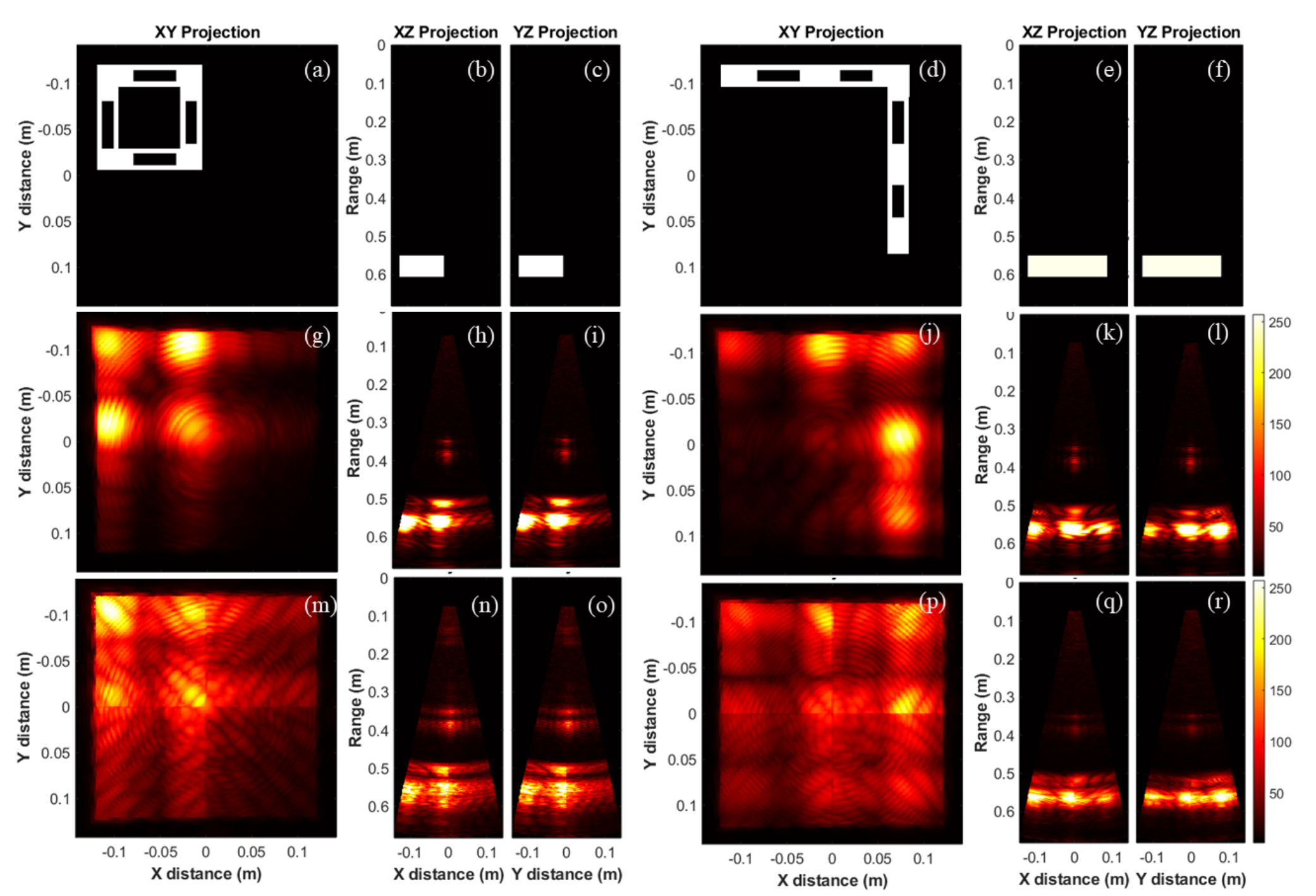}
\caption{$XY$, $XZ$ and $YZ$ projections of the 3D images of Square and L shape targets. (a)-(f) k-Wave Model. (g)-(l) Conventional Method. (m)-(r) Proposed method.}
\label{fig7}
\end{figure*}
\begin{figure*}
\centering
\includegraphics[width=0.85\textwidth]{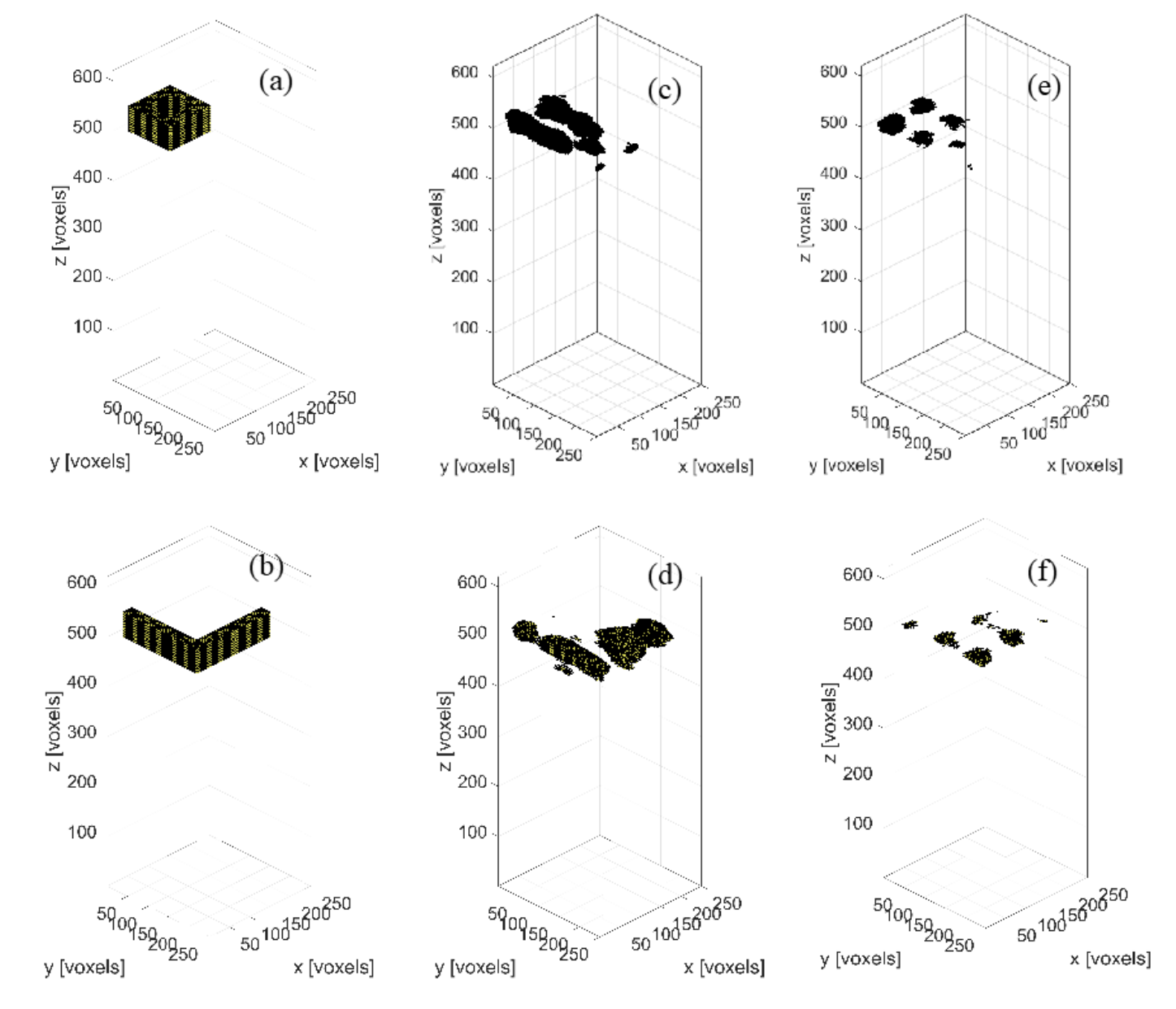}
\caption{ Modeled and reconstructed 3D images. (a)-(b) Modeled targets in the k-Wave grid. 
Gaps within the modeled targets are not detectable in the 3D images due to the limitations imposed by the image size. (c)-(d) Reconstructed 3D images by the conventional method. The gaps within the targets are not visible because of the large MLW of the conventional method.  (e)-(f) Reconstructed 3D images by the proposed method. The gaps within the targets are identifiable because of the reduced MLW.}
\label{fig8}
\end{figure*}
\begin{figure*}
\centering
\includegraphics[width=\textwidth]{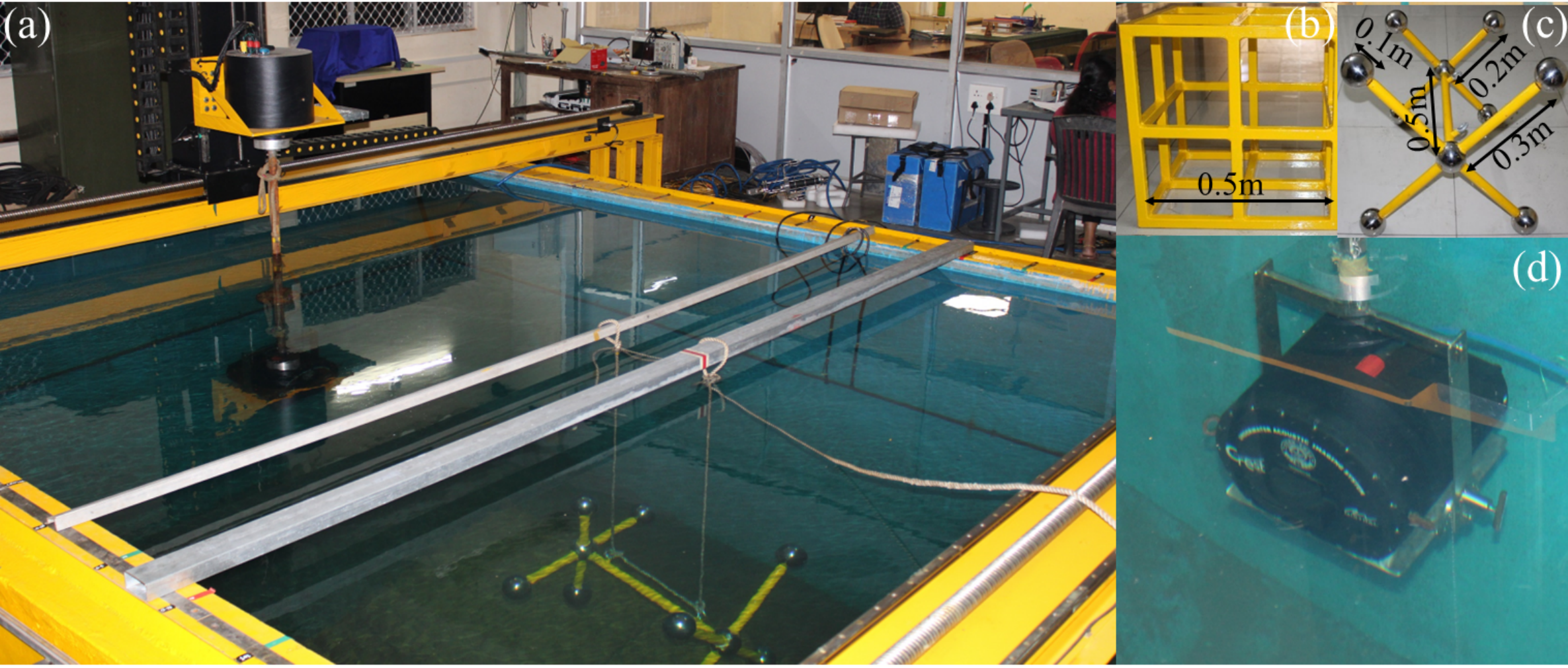}
\caption{Experimental setup. (a) Acoustic tank facility with precise moving system.  (b) Shape and size of the cube shaped target. (c) Shape and size of the target made up of hollow steel balls and cylinders. (d) Sonar system having uniform linear receiving array and omni-directional projector. } 
\label{fig9}
\end{figure*}
\section{UNDERWATER TARGET MODELING AND SIMULATION}
\subsection{Parameters for Underwater Acoustical 3D Imaging}
An underwater acoustical 3D imaging system is designed based on angular resolution and range resolution. Angular resolution is the smallest angle between two objects at the same range that can be distinguished separately. The angular resolution $\theta_{res}$ is expressed as \cite{chi2016ultrawideband}
\begin{equation}
\begin{gathered}
\theta_{\text {res }}=\frac{\lambda}{D} \times \frac{180^0}{\pi} \\
\approx \frac{60 \lambda}{D},
\label{7}
\end{gathered}
\end{equation}
in this context, $\lambda$ represents the wavelength, and  $D$ denotes the array width. It should be noted that, for a given wavelength, an increase in aperture size leads to an improvement in angular resolution. The along track resolution associated with $\theta_{res}$ is given by
\begin{equation}
r_{\text {along track }}=r \theta_{\text {res }},
\label{8}
\end{equation}
where, $r$ represents the imaging range. The along-track resolution varies for all ranges, worsening as the imaging range expands. The range resolution is defined as the smallest distance between two objects in the range direction that can be distinguished separately. It is determined by 
\begin{equation}
r_{\text {range }}=\frac{c}{\Delta f},
\label{9}
\end{equation}
where, $c$ is the acoustic speed and $\Delta f$ is the signal bandwidth.

\begin{figure*}
\centering
\includegraphics[width=\textwidth]{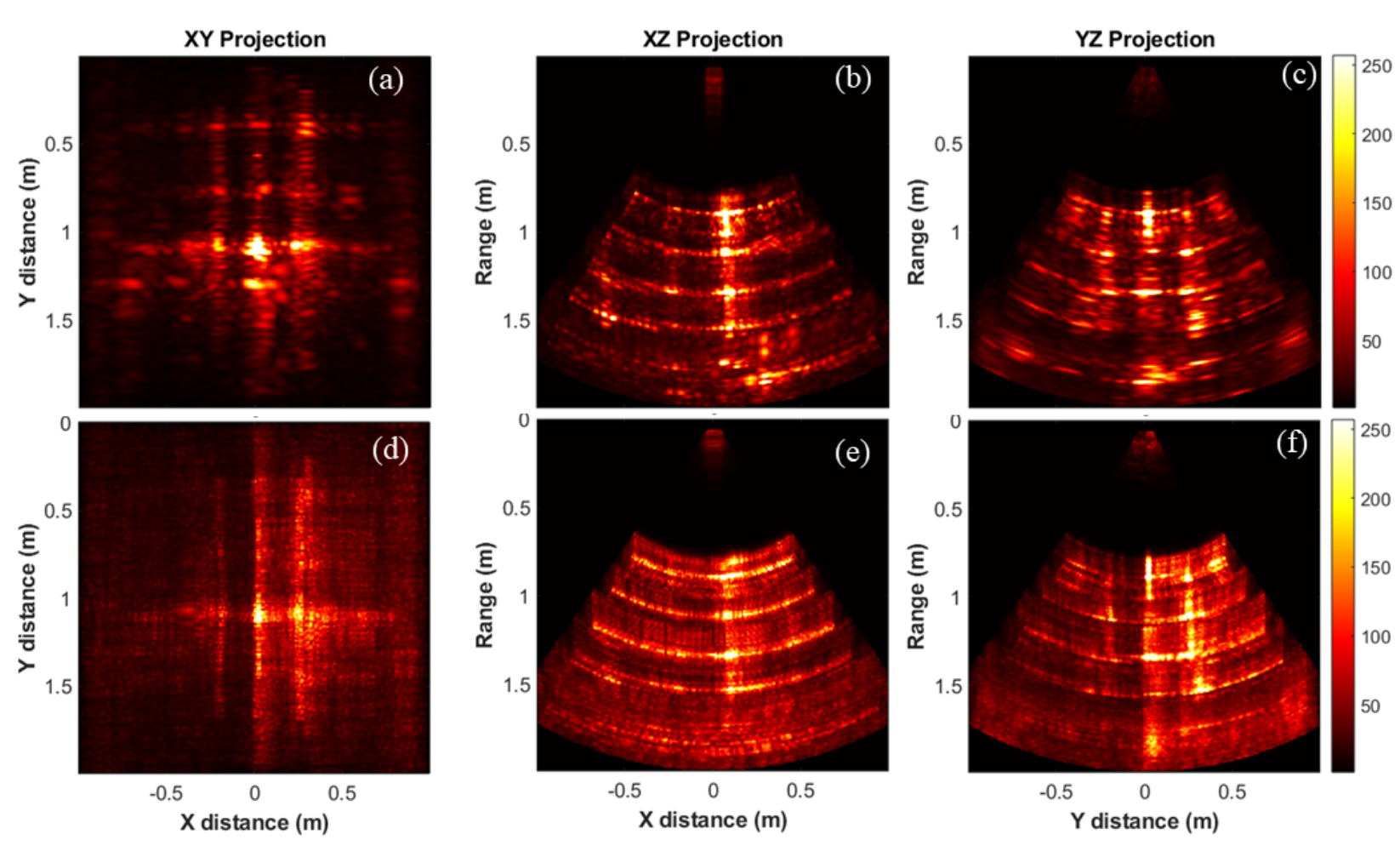}
\caption{$XY$, $XZ$ and $YZ$ projection of the 3D image of the cube shaped target. (a)-(c) The conventional method.  (d)-(f) The proposed method.   }
\label{fig10}
\end{figure*}
\begin{figure*}
\centering
\includegraphics[width=\textwidth]{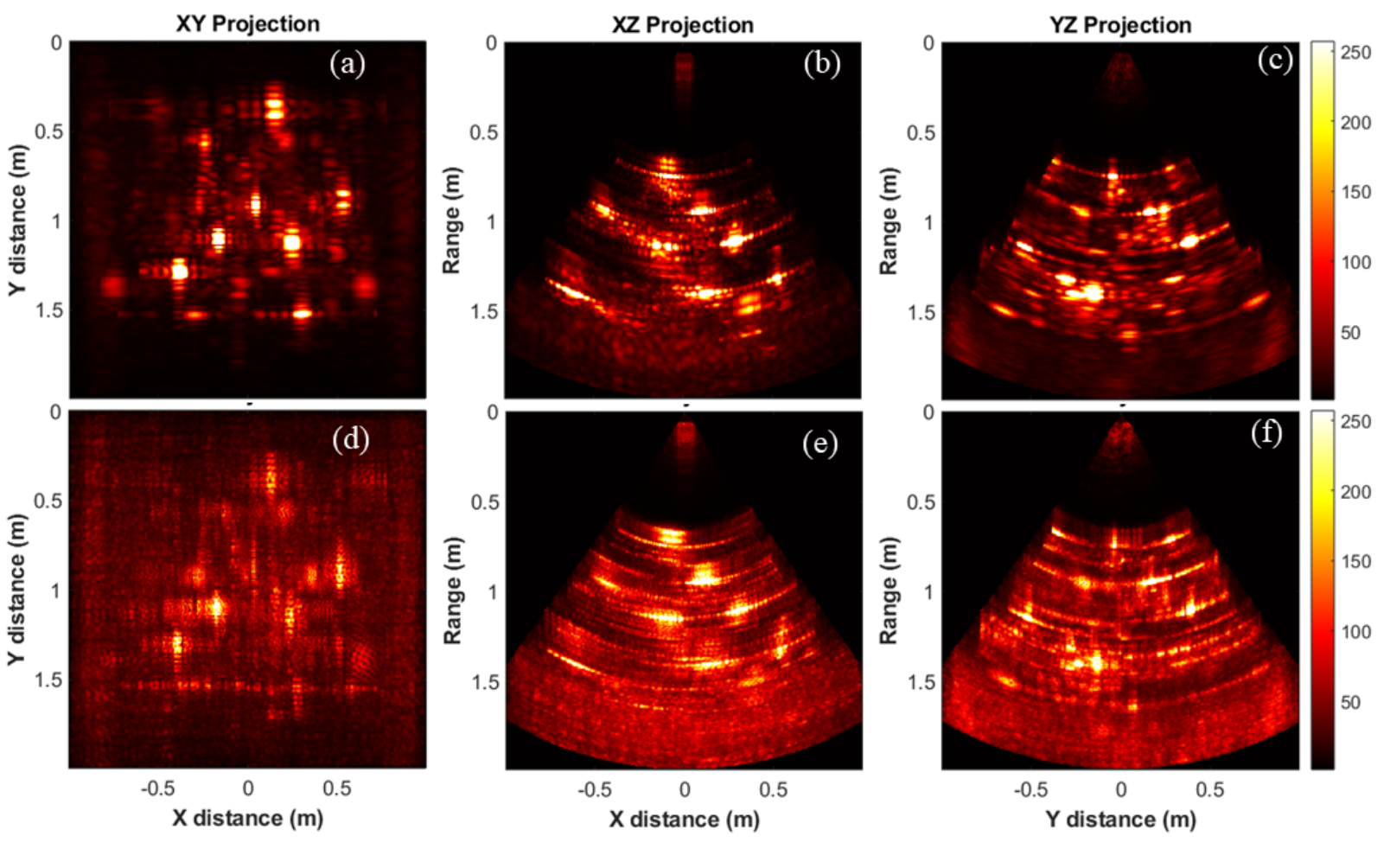}
\caption{$XY$, $XZ$ and $YZ$ projection of the 3D image of the target made up of hollow steel balls and cylinders. (a)-(c) The conventional method.  (d)-(f) The proposed method.   }
\label{fig11}
\end{figure*}
\subsection{Underwater Simulation Using k-Wave}
To verify the developed algorithms for sonar imaging, conducting experimental studies is expensive. Hence, simple analytical modeling techniques were employed \cite{han2013real, zhao2018efficient, qin20183d,wei2019obtaining}. However, these analytical models cannot model effects such as attenuation, reverberation and speckle noise. The k-Wave \cite{treeby2012k} acoustic toolbox provides a more precise solution to the wave equation and becomes a suitable toolbox for underwater target modeling \cite{malekshahi2022solution}. To the best of author's knowledge, the k-Wave simulator has not been looked into for underwater 3D image simulations, despite the fact that it accurately mimics acoustic wave propagation and has been occasionally utilized to simulate underwater environments for diverse applications \cite{satish2021passive}. A previous attempt in this direction was by us in \cite{menakath2022k} where we had modeled underwater targets by lowering the time and memory requirements with a scaled down approach. In this work, we have also extended the approach to 3D underwater target modeling.

For modeling the underwater imaging setup using the k-Wave toolbox, a grid of size $0.6$ $m$$\times$$0.25$ $m$$\times$$0.25$ $m$ was defined\cite{menakath2022k}. The water medium, which contains random scatterers, and the target were characterized within the grid by specifying their density and acoustic speed. A random scatterer distribution with unit mean and $0.01$ standard deviation is adopted.  Mild steel objects of two different shapes, a square and an L shape, were modeled and placed at a distance of $0.5$ $m$ from the receiving array. The shapes are made discontinuous to induce non-surface reflections, which could help in reconstructing the actual shape. Since the target is in far-field, $0.5$ $m$ range was made equivalent to a range of $2.5$  $m$ by defining the 2D uniform sensor array of dimension $48$ $mm$$\times$$48$ $mm$ which provides the same along track resolution of $3.27$ $cm$ for an angular resolution of $0.75^0$ at $2.5$ $m$ range. The angular resolution corresponding to the required along track resolution at $0.5$ $m$ can be calculated using \eqref{8} and the aperture size to achieve the angular resolution can be calculated using \eqref{7}. The sensor array is located at the center of the $xy$ plane within the grid. To avoid ambiguities caused by the grating lobes, an element spacing less than half the wavelength, $1$ $mm$ was used. A single bowl of radius and aperture length, $20$ $mm$ placed in front of the array, is defined as the source. The transmission of wideband signals is ensured by constraining the transmitted pulse to a frequency of $500$ $kHz$ with 3 cycles, resulting in a bandwidth of $218$ $kHz$ and a range resolution of $3$ $mm$.

\section{Simulation results}
For detailed analysis of 3D images, they are projected to different planes by taking maximum of the direction which is omitted. Different projections on different planes of the modeled and reconstructed 3D images of the square and L-shaped targets are shown in Fig.\ref{fig7}. The resulting image from the proposed approach is comparable to that of the conventional method and the reference image of the modeled target. Both methods reconstruct the target image precisely at the same location in the grid where it was modeled. Since the transmitted signal wavelength is larger than the thickness of the strips of the target in Fig.\ref{fig7}(a) and (d), they are not visible in the images in Fig.\ref{fig7}(g),(m),(j) and (p).  Due to the narrow MLW of the proposed method compared to the conventional method, the resolution of the images has been improved by the proposed method. However, background noise has been increased in the proposed method because of the higher level of the side lobe. The modeled and reconstructed 3D images of the underwater target in k-Wave grid are shown in Fig.\ref{fig8}. It should be noted that the modeled targets are not rigid and there exist gaps in between the corners which are not identifiable in Fig.\ref{fig8}(a)-(b). The 3D images reconstructed by the proposed method shows the gaps within the target as shown in Figs.\ref{fig8}(e)-(f). As the MLW of the conventional method is larger than that of the proposed method, the reconstructed images failed to show all the gaps in the targets as shown in Fig.\ref{fig8}(c)-(d). 

\begin{table}
\centering
\begin{center}
\caption{The Table of Comparison}
\label{table2}
\setlength{\tabcolsep}{5pt}
\begin{tabular}{p{70pt}p{35pt}p{35pt}p{35pt}p{25pt}}
\hline\hline
Parameters& 
DAS& 
CZT &
DM&
Proposed Method\\
\hline
Number of real operations& 
$6221952$&
$3.5565\times{10^{11}}$&
$1.9988\times{10^{10}}$&
$864000$\\
Average time (seconds)& 
$0.1210$&
$1.0630$&
$96.2032 $&
$0.0052 $
\\
\hline\hline
\end{tabular}
\label{tab2}
\end{center}
\end{table}

The number of real operations required for $60\times 60$ beams reconstruction using different methods is calculated based on section III.C. and is tabulated in TABLE \ref{table2}. The result shows significant reduction in number of operations for the proposed method as compared to the three existing methods. The computation time required for the reconstruction of a single slice having $60\times 60$ beams is also reported in TABLE \ref{table2}. All the computations are carried out in a system having $11^{th}$ Gen Intel(R) Core(TM) i7-1165G7 @ 2.80GHz, 2803 MHz, 4 Core(s), 8 Logical Processor(s). The results show that the proposed method is $23$ times faster than conventional time domain delay and sum beamforming and $204$ times faster than Chirp Zeta Transform beamforming. It should be noted that the direct implementation of the frequency domain delay and sum is not feasible for wideband signals.

\section{Experimental Results}

Since planar array was unavailable and expensive, we acquired the experimental data by moving a uniform linear array of $96$ elements precisely with a distance equal to the element spacing, $d$ vertically in a tank facility at the Naval Physical and Oceanographic Laboratory (NPOL) as shown in Fig.\ref{fig9}(a). The shape and size of the targets used for the experiment is shown in Fig.\ref{fig9}(b) and (c). The sonar system having a uniform linear array as receiver and an omni-directional projector as transmitter, used for the experiment is shown in Fig.\ref{fig9}(d).  For the experiment, we preserved the tank as static after placing a target and acquired the data $96$ times by placing the sonar system in different heights. We stacked the data as a planar array data and processed using the conventional method as well as the proposed method. The experiment was repeated for two different targets in Fig.\ref{fig9}(b) and (c).  For both experiments, the target is placed at a range of $0.6$ $m$ from the linear array. The initial depth of the sonar system and the depth of the target is $1.5$ $m$ from the surface of the water.

The experimental results obtained from a cube shaped target is shown in Fig.\ref{fig10}. The target is placed in the tank in such a way that one face of the cube is parallel to the receiving array. The video captured by rotating the segmented 3D image of the target obtained by the conventional method and the proposed method is presented in the multimedia material, MM1.Rotation video of the cube target. Since high-intensity reflections are coming from the corners of the cube, they are clearly visible in the reconstructed image as shown in the video and in Fig.\ref{fig10}. The proposed method has reconstructed the 3D image of the target at the same position and almost in the same shape and size as shown in the video. Since the side lobe level of the proposed method is higher than that of conventional method, noise in the background has been increased. However, the proposed method reconstructed the 3D image in $31.3628$ seconds, while the conventional method reconstructed it in $3.0560\times 10^3$ seconds using the system having a 2* Intel Xenon Gold 6248 2.5 GHz processor, 128 GB RAM and Nvidia Quadro RTX4000, 8GB, 3DP, VirtualLink(XX20T) display card. For detailed analysis, projection of the 3D image on different planes for the conventional method and the proposed method is presented in Fig.\ref{fig10}(a)-(c) and (d)-(f) respectively. The projection is done by taking the maximum of the dimension that is removed. While comparing the projected images, it is clear that the proposed method has captured all the high intensity reflections as in the conventional method with some background noise. The effect of background noise can be avoided by proper thresholding and proper dynamic range selection for the display. For comparison, we have taken the same dynamic range for the conventional method and the proposed method regardless of the image clarity.

The experimental results obtained from a target made up of hollow steel balls and rods are shown in Fig.\ref{fig11}. The target is placed in the tank in such a way that smaller face of the target is near to the receiving array with an inclination. The video captured by rotating the segmented 3D image of the target made up of hollow steel balls and cylinders obtained by the conventional method and the proposed method is presented in the multimedia material, MM2.Rotation video of the the target made up of hollow steel balls and cylinders. Since high intensity reflections are coming from the balls, they are clearly visible in the video as well as in Fig.\ref{fig11}. The proposed method has reconstructed the 3D image of the target at the same position with the same inclination, shape and size as in the conventional method. The proposed method reconstructed the 3D image in $35.9942$ seconds, while the conventional method reconstructed it in $3.5235\times 10^3$ seconds using the same processing system.  Projection of the 3D image on different planes for the conventional method and the proposed method is presented in Fig.\ref{fig11}(a)-(c) and (d)-(f) respectively. The same observation as in the previous case is shown in Fig.\ref{fig11}. High intensity reflections have been highlighted in the proposed method with some background noise due to side lobes compared to the conventional method.

\section{Conclusion}
The paper presented a novel quadrant-based non-linear beamforming algorithm using an L-shaped  array (placed at the edges of a large rectangle array) in the time domain for fast implementation of acoustical 3D imaging. The proposed method demonstrated a reduction in computational complexity and computation time over the three state-of-the-art methods with ${1}^\circ$ degree improvement in angular resolution and $11.8$ $dB$ increase in the peak side lobe level. Although there is an increase in the background noise due to side lobes, the same is restricted to a quadrant using the method and could be thresholded appropriately. A through analysis of the proposed L-shaped array for acoustical 3D imaging in the far field is done using the point spread function. The 3D images of two different underwater targets modeled in k-Wave were reconstructed using the proposed approach without degrading the image quality. The experimental results of two different underwater targets also demonstrated the same observation as in simulation with a reduction in the computation time by a factor of $97$.

\section*{acknowledgments}
The authors would like to acknowledge the funding from the Defense Research and Development Organization (DRDO) and the Ministry of Education (MoE), India. Authors would also like to acknowledge the high-performance computational facility at IIT Palakkad and at NPOL Kochi in accelerating the simulation and beamforming studies. The authors express their gratitude to all those who assisted with the experiments conducted at the acoustic tank facility at NPOL Kochi.

\bibliographystyle{ieeetr}
\bibliography{Reference}
\end{document}